# Air Shower Radio Emission with Energy $E_0 \geq 10^{19}$ eV by Yakutsk Array Data.

S. P. Knurenko and I. S. Petrov

Yu. G. Shafer Institute of Cosmophysical Research and Aeronomy SB RAS, Yakutsk, Russia.

The paper presents short technical description of Yakutsk Radio Array and some preliminary results obtained from measurements of radio emission at 32 MHz frequency induced by air shower particles with energy $\varepsilon \geq 1 \cdot 10^{19}$ eV. The data obtained at Yakutsk array in 1987-1989 (first set of measurements) and 2009-2014 (new set of measurements). For the first time, at Yakutsk array radio emission from air shower with energy $> 10^{19}$ eV was registered including the shower with highest energy ever registered at Yakutsk array with energy $\sim 2 \cdot 10^{20}$ eV.

## 1. Introduction

The method of registration of radio emission of ultrahigh-energy particles is based on Askaryan effect. The effect proposes that particle shower develops a negative charge excess, accompanying the passage of ultra-high energy particles through matter [1]. According to this effect, excess of electrons in showers is caused by the annihilation of positrons with electrons of the Earth's atmosphere. The movement of the shower disc with negative charge excess at a speed greater than the phase velocity of light in that medium is the cause of Cherenkov radiation emission at radio frequencies [2]. In the following years after this discovery, there have been many experimental studies of radio emission from air showers [3, 4], including the Yakutsk array [5]. Short reviews of the air shower radio emission work can be found in [4, 6, 7]. In paper [7] pointed out the possibility of registration of air showers with energies above $10^{19}$ eV, employing radio equipment placed on the surface of the Earth and registering the radio emission by satellites on the Earth orbit. Surface arrays require a huge area of 3-5 thousand square kilometers for the registration of showers with such energies. In addition, it requires relatively quiet in terms of radio interference place in the urbanized society that is difficult to find. At the same time, satellite based arrays would allow a large solid angle which covers bigger areas and detect a larger number of air showers with highest energies. Thus, the problem of statistics of such showers would have been solved, and the spectrum of cosmic rays would be studied at energies up to $10^{21}$ eV. However, before one put this idea into practice, we need to ensure the effectiveness of this method of registration of showers with ultra-high energies. For these purposes would most suited the currently existing large ground arrays where exist a corresponding infrastructure which can be used for registration of radio emission. Experiments on radio radiation from the EAS were actively carried out in 60 - 70 years of the last century. For example, the array of the Moscow State University in the 70s registered air shower radio emission at energies $10^{16}$ - $10^{17}$ eV [8, 9]. Later, in 1986 - 1989, at the Yakutsk array were carried out measurements of radio emission in energy range above $10^{17}$ eV [5, 10, 11].

In recent years, interest in the air shower radio emission, as an independent method to study the physics of the EAS has grown significantly, and for registration of radio emission were built arrays of significant size [12, 13]. This method makes it possible not only to evaluate the energy, but also to reconstruct the longitudinal shower development, namely, the depth of maximum $X_{max}$ [14, 15]. This is especially important for huge arrays where the uncertainty in the estimation of shower energy with different methods of detecting air showers reaches about (20-40)%. For example, Auger and Telescope Array difference is 20% and the cause of differences is still remains unknown [16]. Thus, the radio emission, in conjunction with other methods of could be employed for intercalibration of huge arrays.

This paper presents radio emission of EAS with ultra-high energies data obtained by Yakutsk array in 1987-1989 and 2009-2014 years.

## 2. Observation of Air Showers Radio Emission at Ultra-High Energies.

### 2.1. Radio array.

In the mid 80-es of the last century, the Radio Array with registration bandwidth 30-40 MHz was designed as an extension of main Yakutsk particle array [5]. The setup consisted of two parts: analog and digital. The analog part comprises the reception, amplification of the radio signal, matching circuits of the output signals by the level and frequency with the parameters of the digital recorder. The digital part of the array converts input analog signals into digital code and writes information of the radio noise state and the signal from the shower to a buffer RAM. Then the information about the noise field and a radio pulse from EAS were copied to the computer hard disc drive (HDD). Twenty receiving antennas, which are installed on 10 pillars as shown in Figure 1, register air shower radio emission.

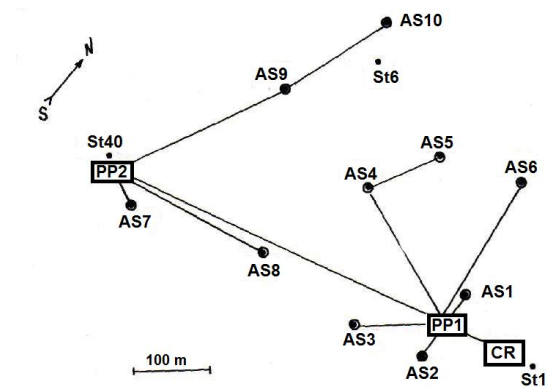

Fig. 1. Arrangement of the radio antennas, in 1987 - 1989. AS - antenna station; PP - peripheral (intermediate) collection point for air shower data; CR - central registration point; ST – station with scintillation detectors of Yakutsk array.

The distance between antenna pillars was 50, 100, 200, 300 and 500 meters, covering an area of roughly 0.35 km$^2$. One pillar consists two independent half-wave dipoles with orientation E-W and N-S. Antennas installed at $\lambda/4$ above ground, thus ensuring a maximum of the radiation pattern for the emission coming from the top.

Figure 2 shows the location of antennas. At the lower part of antenna, special container is located. In order to enhance the radio signals, unified broadband receivers with direct amplification at bandwidth of 30-35 MHz were applied. Suppression of the gain at frequency 29 MHz $\geq$ 40 dB, at frequencies less than 28 MHz $\geq$ 60 dB.

Constructively, the receivers are designed as two blocks. The first block is located under the antenna, the block consists a low noise amplifier with a gain of $Ku \cong 40$ dB and output matching with the cable. In the second block, final amplifier with gain $Ku \cong 40$ dB was placed. To match with a bandwidth of the ADC at the output of the amplifier amplitude detectors are used. To improve detection of linearity powerful FET (Field Effect Transistor) type KP901 and KP902 were used.

All recording equipment: power amplifiers, detectors and ADC were placed in the two warm cabins, because of the extremely low winter temperatures (-40 ° C). Also the cabins contains calibration generators and high-frequency switches.

## 2.2. Calibration

During calibration process, the input of the antenna amplifier is disconnected from the antenna and is connected to the output of the calibration generator via coaxial cable. Calibration is performed automatically without operator intervention at specified intervals of time. For this purpose, the remote-controlled generators G4-151 and RF switches on the relay of REV-15, which is controlled by a central computer, were used. To improve the accuracy of timing synchronization of additional ADC of crystal oscillators has been introduced.

## 2.3. ADC

In the first stage of the experiment ADC F-4226 with the following parameters were used: conversion time - 50 ns, accuracy - 8 bits (256 amplitude points), RAM-1024 word capacity (51 ms). Continuous operation of the converter allows one to store information in memory of the radio pulses before receiving the ADC trigger signal input from the "master" of the main Yakutsk array. 9th bit of data word is a sign of the data are in the RAM to run.

Additional synchronization with EAS provided by a separate channel of signal detection for time synchronization at a frequency of 207 MHz with an accuracy of 100 ns, using the same type of ADC.

# 3.Results

In 10 years of radio array operation, observation time were 50400 hours. 14700 showers with axis within a circle of 1.5 km and with energy $\geq 10^{17}$ eV were registered by radio array. In addition, there were a few showers with energy $\geq 10^{19}$ eV. From these data, we found a correlation between radio emission and air shower parameters [15] and proved that it is possible to study physics and astrophysics of cosmic rays (CR) from only radio emission data [16].

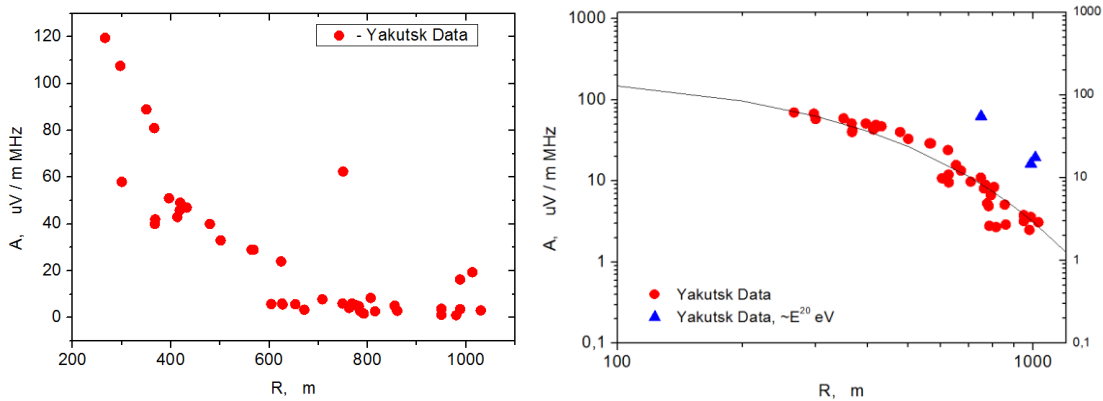

Fig. 3. Dependence of LDF of electric intensity $\varepsilon_v$ from air shower axis. Dots are normalized to mean energy $<E_0>$ = 1.54·10$^{19}$ eV. a) Decimal scale b) Logarithmic scale. Triangles – signals of air shower with energy ~10$^{20}$ eV.

Fig. 3a shows cloud of dots that plotted by data listed in table 1. Most of the showers have energy from $10^{19}$ to $3.5 \cdot 10^{19}$ eV, and two showers with energy $\geq 10^{20}$ eV. In Fig. 3b dots are normalized to mean energy $<E_0>$ = 1.54·10$^{19}$ eV and mean zenith angle $<\theta>$ = 43.1° and in logarithmic scale.

Table 1. List of air showers with energy 10¹⁹ eV registered by Yakutsk array antennas. date – is a date of shower registration, θ – zenith angle (degree), ψ – azimuth angle (degree), $E_0$ – energy of primary particle (eV), $A_v$ – radio emission amplitude (μV·m⁻¹·MHz⁻¹), R – distance from air shower axis to antenna (m).

| date | $\theta$, deg | $\psi$, deg | $E_0$, эB | $A_\nu$ | R, m |
|---|---|---|---|---|---|
| 16.11.86 | 74 | 180 | $3.1 \cdot 10^{19}$ | 58.0 | 300 |
| 16.12.87 | 71 | 178 | $3 \cdot 10^{19}$ | 40 | 367 |
| 21.02.88 | 70 | 210 | $10^{19}$ | 3.1, 3.8 | 1030, 950 |
| 09.03.88 | 36 | 125 | $9 \cdot 10^{18}$ | 6.2 | 792 |
| 07.05.89 | 59 | 168 | $2 \cdot 10^{20}$ | 62.5 | 750 |
| 10.03.11 | 51 | 239 | $1.1 \cdot 10^{19}$ | 89, 43, 5.8 | 350, 413, 604 |
| 16.05.11 | 69 | 99 | $1.6 \cdot 10^{19}$ | 33, 29, 40 | 501, 564, 479 |
| 31.12.11 | 15 | 165 | $1.1 \cdot 10^{19}$ | 1.2, 1.0, 2.9 | 950, 980, 860 |
| 12.04.12 | 8 | 222 | $1.3 \cdot 10^{19}$ | 4.1, 2.8, 6.0 | 762, 785, 626 |
| 04.05.13 | 46 | 295 | $1.1 \cdot 10^{19}$ | 5.3, 6.0, 12 | 776, 768, 368 |
| 12.12.13 | 15 | 297 | $1.2 \cdot 10^{19}$ | 5.1, 8.4, 3.6 | 855, 806, 988 |
| 03.10.13 | 21 | 21 | $1.1 \cdot 10^{19}$ | 9.1, 11, 2.7 | 419, 396, 815 |
| 22.03.13 | 46 | 4 | $1.8 \cdot 10^{19}$ | 41, 48, 78 | 418, 432, 366 |
| 02.01.14 | 48 | 207 | $1.0 \cdot 10^{20}$ | 16.3, 19.4 | 1013, 988 |
| 22.01.14 | 47 | 189 | $1.1 \cdot 10^{19}$ | 107.6, 119.6 | 297, 266 |
| 05.02.14 | 26 | 343 | $3.5 \cdot 10^{19}$ | 3.4, 5.6 | 671, 627 |
| 02.03.14 | 30 | 217 | $1.2 \cdot 10^{19}$ | 4.9, 6.0, 7.8 | 782, 749, 708 |

In Fig. 3b there are triangles that significantly higher than other dots. These signals are registered in two showers with maximum energy. We didn't normalize these energy to emphasize the special status of these points, because they belong to the showers with energy $E_0 \geq 10^{20}$ eV and their amplitude greater than in other showers. Another shower with energy $10^{20}$ eV were registered in 7 May 1989. A second shower (4 January 2014) was able to register by two antennas at different distances (1013 m and 988 m from the shower axis).

We see that after normalization cloud of dots indicates rapid attenuation of radio signal and shows lateral distribution function (LDF) dependence on the distance of antennas from air shower axis. As can be seen from Fig. 3a the resulting dependence of the amplitude of radio signal from distance close to exponential function:

$$E = \varepsilon \cdot \exp\left(\frac{R}{R_o}\right) \quad (1)$$

In order to describe experimental data we used minimization method and found a function:

$$\varepsilon = (29,5 \pm 1,6)\left(\frac{E}{5 \cdot 10^{17}}\right)^{(0,83 \pm 0,03)} \cdot (1 - \cos\theta)^{1,16 \pm 0,03} \cdot \exp\left\{-\frac{R}{[(162 \pm 8) + (84 \pm 3)\left[\frac{(X - 675)}{100}\right]]}\right\}, \quad (2)$$

where $E_0$ – energy of primary particle, $\theta$ – zenith angle, R – distance from antenna to air shower axis, $X_{max}$ - depth of maximum development.

Formula (2) well describes LDF of radio emission at medium and large distances from air shower. In Fig. 3b approximation is show by solid line.

Using this, we determined amplitude of the signal at a distance of 175 m and 725 m and the calculated the depth of shower maximum $X_{max}$ by the formula (3):

$$X_{max} = (660 \pm 15) + (100 \pm 5)\left(\frac{P-11.5}{3}\right), \quad (3)$$

where $P$ – a parameter that characterize the ratio of amplitude of radio emission at selected distances from air shower axis, $P = A_1/A_2$.

By applying connection between $X_{max}$ and slope value that found in [15], we estimated the average $X_{max}$ = 760±30 g·cm$^{-2}$ for the hypothetical shower with energy 1.54·10$^{19}$ eV.

Now, we use experimentally obtained $X^{exp}_{max}$ to estimate atomic mass of primary particle of cosmic ray $<lnA>$, that produced air shower with that energy. One can do that by interpolation method (formula (4))[17]:

$$<\ln A> = \left(\frac{X^{exp}_{max} - X^{p}_{max}}{X^{Fe}_{max} - X^{p}_{max}}\right) \cdot \ln A_{Fe}, \quad (4)$$

where $X^{exp}_{max}$ – depth of maximum development determined by radio data, $X^{p}_{max}$ – calculated depth maximum of primary proton, $X^{Fe}_{max}$ – calculated depth of maximum of primary iron nucleus. $lnA_{Fe}$ - atomic mass of iron. Depth of maximum development were made by QGSJettII-04 hadron interaction model.

Resulting value of $<lnA>$ = (1.61±0.4), is very close to $<lnA>$ = (0.8 – 2.0) obtained by Cherenkov light data [18]. In this case, we can say that the flow of primary CR particles that form air shower with energies above 10$^{19}$ eV, consists mainly of protons p, helium nuclei He and carbon nuclei C.

Thus using radio method in the highest energies region it is possible to obtain primary characteristics of air shower: energy E, air shower power N and depth of maximum $X_{max}$ like in other ground detection methods (scintillation detectors and detectors of Cherenkov light).
Radio array with bigger area will have higher statistics of air showers with energy ≥10$^{19}$ eV. It will allow obtaining astrophysics characteristics of CR like energy spectrum and mass composition (MC) of primary particles.

## 4. Conclusion

Long-term observation of air shower radio emission at Yakutsk proved existence of radio emission at energies ≥10$^{19}$ eV. It allowed us to determine some characteristics of air showers at these energies.

Particularly, a function of air shower radio emission attenuation at energies ≥10$^{19}$ eV was obtained and its gradient was estimated. Further, an existence of radio emission at energies ≥10$^{20}$ eV, i.e. air showers with highest energies registered at Yakutsk array [8] was proved. Also, there is a significant signal in highly inclined showers and effect in LDF shape from geomagnetic field. Latter once again indicates to the geomagnetic generation of radio emission [14]. Lastly, in the frame of QGSJETII-04 model, estimation of CR mass composition at energies 10$^{19}$ eV obtained by radio data in a good agreement with results of [18]. MC indicates to CR primarily consisting light nuclei. This conclusion doesn't contradict to the results of MC obtained by huge air shower ground arrays [19].

## Acknowledgment.

Authors express their gratitude to Yakutsk array collaboration for use of experimental data and useful discussions during preparation of the article. In addition, authors thanks the reviewers for